\documentclass[12pt]{article}
\usepackage{latexsym,amsfonts,amssymb}
\makeatletter
\@addtoreset{equation}{subsection}
\makeatother

\begin{document}
\begin{center}
{\Large \bf On Dark Matter Problem:\\ Pseudomatter---Concept and
Applications}
\\[1.5cm]
 {\bf Vladimir S.~MASHKEVICH}\footnote {E-mail:
  Vladimir.Mashkevich100@qc.cuny.edu}
\\[1.4cm] {\it Physics Department
 \\ Queens College\\ The City University of New York\\
 65-30 Kissena Boulevard\\ Flushing, New York
 11367-1519} \\[1.4cm] \vskip 1cm

{\large \bf Abstract}
\end{center}

A solution to the dark matter problem is set forth in the
framework of reductive semiclassical gravity, i.e., semiclassical
gravity involving quantum state reduction. In that theory, the
Einstein equation includes the energy-momentum tensor originating
from pseudomatter and partially compensating for quantum jumps of
the matter energy-momentum tensor. The compensation ensures the
continuity of metric and of its first time derivative.
Pseudomatter is actualized as pseudodust and perceived as a dark
matter. The necessity of compensating for quantum jumps makes
pseudomatter, i.e., dark matter of such a form, an indispensable
rather than ad hoc element of the theory. Applications: The
Schwarzschild solution with pseudomatter, pseudomatter halo,
collapse involving pseudomatter, pseudomatter in the FLRW
universe.

\newpage

\section*{Introduction}

It is commonly known that astrophysics and cosmology are being
confronted with the problem of dark matter [1--6]. In the
framework of general relativity, which is a totally classical
theory, the problem is a riddle: dark matter appears to be an
entirely arbitrary element---what does it exist for?

Meanwhile, reductive semiclassical gravity, i.e., semiclassical
gravity incorporating quantum state reduction involves an
indispensable element, namely, pseudomatter, which is actualized
as pseudodust and perceived as a dark matter [7].

In reductive semiclassical gravity, the Einstein equation includes
the energy-momentum tensor originating from pseudomatter. That
tensor partially compensates for jumps of the matter
energy-momentum tensor resulting from state quantum jumps. The
compensation ensures the continuity of metric and of its first
time derivative, which is necessary for the Einstein equation to
be fulfilled. It cannot be too highly stressed that in reductive
semiclassical gravity dark matter in the form of pseudomatter is
an indispensable rather than ad hoc element of the theory.

The idea of introducing a nonmaterial field into the Einstein
equation had been advanced in [8]. The idea was realized in the
form of pseudomatter in [9] and further developed in [10,7].

In this paper, the concept of pseudomatter is elaborated in
detail.

Some applications of the concept to astrophysics and cosmology are
given: the Schwarzschild solution with a realistic matter state
equation due to pseudomatter, pseudomatter halo (an important
result: the velocity $v=\mathrm{const}$), collapse involving
pseudomatter, pseudomatter in the FLRW universe (an essential
result: the pseudomatter density $\varepsilon \propto 1/R$ where
$R$ is the radius of the universe).

\section{Fundamentals: Reductive semiclassical gravity}

\subsection{Semiclassical gravity}

Semiclassical gravity is based on the semiclassical Einstein
equation and the Schr\"odinger equation.

For what follows, it is convenient to introduce the tensor
\begin{equation}\label{1.1.1}
 E_{\mu\nu}=G_{\mu\nu}-\Lambda
g_{\mu\nu}-8\pi \varkappa T_{\mu\nu}\,,\quad \mu,\nu=0,1,2,3
\end{equation}
where $G$ is the Einstein tensor, $\Lambda$ is the cosmological
constant, $\varkappa$ is the gravitational constant, $g$ is
metric, $T$ is the energy-momentum tensor; $c=1,\;\hbar=1$, so
that $\varkappa=T_{\mathrm{P}}^{2}$ ($T_{\mathrm{P}}$ is the
Planck time). $E$ may be called the dynamical Einstein tensor.
Next,
\begin{equation}\label{1.1.2}
T_{\mu\nu}=(\Psi,\hat{T}_{\mu\nu}\Psi)
\end{equation}
where $\hat{T}_{\mu\nu}$ is the energy-momentum tensor operator
and $\Psi$ is a state vector of quantized matter, i.e., quantum
fields $\hat{\phi}$. Thus
\begin{equation}\label{1.1.3}
G_{\mu\nu}=G_{\mu\nu}[g]
\end{equation}
\begin{equation}\label{1.1.4}
\hat{T}_{\mu\nu}=\hat{T}_{\mu\nu}[g,\hat{\phi}]
\end{equation}

The semiclassical Einstein equation is
\begin{equation}\label{1.1.5}
E=0\qquad E_{\mu\nu}=0
\end{equation}

The Schr\"odinger equation is
\begin{equation}\label{1.1.6}
\frac{\mathrm{d}\Psi}{\mathrm{d}t}=-\mathrm{i}\hat{H}\Psi
\end{equation}
where $\hat{H}$ is the Hamiltonian and $t=x^{0}$ is a global time
coordinate.

\subsection{Quantum state reduction and violation of the Einstein equation}
Quantum state reduction is a jump of the state vector:
\begin{equation}\label{1.2.1}
\Psi_{\mathrm{before\,jump}}=:\Psi^{<}
\stackrel{\mathrm{jump}}{\longrightarrow}\Psi^{>}:=\Psi_{\mathrm{after\,jump}}
\end{equation}
The jump of $\Psi$ results in that of $T$:
\begin{equation}\label{1.2.2}
\Delta T=(\Psi^{>},\hat{T}\Psi^{>})-(\Psi^{<},\hat{T}\Psi^{<})
\end{equation}
under the assumption that $\hat{T}$ is continuous. The
discontinuity of $T$ causes a violation of the semiclassical
Einstein equation (1.1.5). Indeed, the latter may be written as
\begin{equation}\label{1.2.3}
G_{ij}-\Lambda g_{ij}-8\pi\varkappa T_{ij}=0\,,\quad i,j=1,2,3
\quad (6\;\mathrm{equations})
\end{equation}
\begin{equation}\label{1.2.4}
G_{0\mu}-\Lambda g_{0\mu}-8\pi\varkappa T_{0\mu}=0\,,\quad
\mu=0,1,2,3 \quad (4\;\mathrm{equations})
\end{equation}
The space components of the Einstein tensor, $G_{ij}$, involve the
second time derivatives $\ddot{g}_{kl}$ of the metric tensor
components $g_{kl}$ [11,12]. It is reasonable to assume that
quantum jumps of $T_{ij}$ result in those of $\ddot{g}_{kl}$. That
is quite conceivable from the physical point of view: A jump of
force ($T_{ij}$) causes a jump of acceleration ($\ddot{g}_{kl}$).
On the other hand, the time/time-space components $G_{0\mu}$
involve only the first time derivatives $\dot{g}_{kl}$ [11,12], so
that jumps of $T_{0\mu}$ do harm: a jump of $\dot{g}_{kl}$ results
in the $\delta$-function in $\ddot{g}_{kl}$, which violates
(1.2.3). (A jump of $g$ would result in the $\delta$-function in
$\dot{g}$.)

\subsection{Compensatory field: Pseudomatter}

In order to compensate for the jumps of $G_{0\mu}$, it is
necessary to extend (1.1.5). Put
\begin{equation}\label{1.3.1}
E_{\mu\nu}=8\pi\varkappa T_{\mathrm{ps}\,\mu\nu}
\end{equation}
where ps stands for pseudomatter, so that
\begin{equation}\label{1.3.2}
G_{\mu\nu}-\Lambda g_{\mu\nu}
=8\pi\varkappa(T_{\mu\nu}+T_{\mathrm{ps}\,\mu\nu})
\end{equation}
The components $G_{0\mu}$ must be continuous, therefore the
components $T_{\mathrm{ps}\,0\mu}$ have to compensate for the
jumps of the four components $T_{0\mu}$. Thus the compensatory
tensor $T_{\mathrm{ps}\,\mu\nu}$ should involve four functions on
the spacetime manifold $M^{4}$.

\subsection{Pseudomatter as pseudodust. The extended Einstein equation}

The most appropriate way of introducing four functions is to treat
pseudomatter as pseudodust:
\begin{equation}\label{1.4.1}
T_{\mathrm{ps}}=\varepsilon\upsilon\otimes\upsilon\qquad
T_{\mathrm{ps}\,\mu\nu}=\varepsilon\upsilon_{\mu}\upsilon_{\nu}
\end{equation}
with
\begin{equation}\label{1.4.2}
\upsilon^{\mu}\upsilon_{\mu}=s(\upsilon)\in\{1,-1,0\}
\end{equation}
It is pertinent to note that the vector $\upsilon$ is not a
four-velocity: for the latter $s(\upsilon)=1$.

Thus the extended Einstein equation is of the form
\begin{equation}\label{1.4.3}
E_{\mu\nu}=8\pi\varkappa\,\varepsilon\upsilon_{\mu}\upsilon_{\nu}
\end{equation}
or
\begin{equation}\label{1.4.4}
G_{\mu\nu}-\Lambda g_{\mu\nu}=8\pi\varkappa(T_{\mu\nu}+
\varepsilon\upsilon_{\mu}\upsilon_{\nu})
\end{equation}

It is obvious that pseudomatter is perceived as a dark matter: it
interacts with matter only via gravity.

There is no intrinsic dynamics of pseudomatter, and
\begin{equation}\label{1.4.5}
T_{\mathrm{ps}}{}^{\mu\nu}{}_{;\nu}=0\quad \mathrm{does\; not\;
hold\; by\;itself}
\end{equation}
so that the equation
\begin{equation}\label{1.4.6}
(T+T_{\mathrm{ps}})^{\mu\nu}{}_{;\nu}=0
\end{equation}
holds only as a trivial consequence of (1.3.2) and
\begin{equation}\label{1.4.7}
(G-\Lambda g)^{\mu\nu}{}_{;\nu}\equiv 0
\end{equation}
Therefore, there are 10 dynamical equations (1.4.3): the equations
\begin{equation}\label{1.4.8}
E_{0\mu}=8\pi\varkappa\,\varepsilon\upsilon_{0}\upsilon_{\mu}
\end{equation}
are dynamical as well as the equations
\begin{equation}\label{1.4.9}
E_{ij}=8\pi\varkappa\,\varepsilon\upsilon_{i}\upsilon_{j}
\end{equation}

\subsection{The reduced Einstein equation}

The pseudomatter can be easily eliminated from the extended
Einstein equation. We have
\begin{equation}\label{1.5.1}
E_{00}E_{\mu\nu}=(8\pi\varkappa)^{2}
\upsilon_{0}\upsilon_{0}\upsilon_{\mu}\upsilon_{\nu}=
E_{0\mu}E_{0\nu}
\end{equation}
Introduce the metrodynamical tensor
\begin{equation}\label{1.5.2}
M_{\mu\nu}=E_{00}E_{\mu\nu}-E_{0\mu}E_{0\nu}\qquad
M_{ij}=E_{00}E_{ij}-E_{0i}E_{0j}\qquad M_{0\mu}\equiv 0
\end{equation}
Thus we obtain the metrodynamical, or the reduced Einstein
equation
\begin{equation}\label{1.5.3}
M_{ij}=0\,,\;\; i,j=1,2,3
\end{equation}
In the six equations (1.5.3), jumps of $T_{\mu\nu}$ are balanced
by those of $\ddot{g}_{kl}$.

The pseudomatter variables are expressed in terms of $E_{\mu\nu}$.
We have
\begin{equation}\label{1.5.4}
\sum_{\mu}E_{\mu\mu}=8\pi\varkappa\,\varepsilon\sum_{\mu}(\upsilon_{\mu})^{2}
\end{equation}
and
\begin{equation}\label{1.5.5}
E_{\mu}^{\mu}=8\pi\varkappa\,\varepsilon s(\upsilon)
\end{equation}

Let
\begin{equation}\label{1.5.6}
\epsilon\upsilon\otimes\upsilon\neq 0\qquad \varepsilon\neq
0\qquad \upsilon\neq 0
\end{equation}
then
\begin{equation}\label{1.5.7}
\sum_{\mu}(\upsilon_{\mu})^{2}>0\qquad \sum_{\mu}E_{\mu\mu}\neq 0
\end{equation}
and
\begin{equation}\label{1.5.8}
\mathrm{sign}\,\varepsilon=
\mathrm{sign}\left(\sum_{\mu}E_{\mu\mu}\right)=\pm 1
\end{equation}
Next,
\begin{equation}\label{1.5.9}
s(\upsilon)=(\mathrm{sign}\,\varepsilon)\mathrm{sign}\,E_{\mu}^{\mu}
\qquad \mathrm{sign}\,E_{\mu}^{\mu}=1,-1,0
\end{equation}

If
\begin{equation}\label{1.5.10}
E_{\mu}^{\mu}\neq 0
\end{equation}
then
\begin{equation}\label{1.5.11}
\varepsilon=\frac{1}{8\pi\varkappa}E_{\mu}^{\mu}s(\upsilon)=
\frac{1}{8\pi\varkappa}(\mathrm{sign}\,\varepsilon)E_{\mu}^{\mu}\,
\mathrm{sign}\,E_{\mu}^{\mu}=\frac{1}{8\pi\varkappa}(\mathrm{sign}\,\varepsilon)
|E_{\mu}^{\mu}|
\end{equation}
and
\begin{equation}\label{1.5.12}
\upsilon_{\mu}\upsilon_{\nu}=\frac{1}{8\pi\varkappa\,\varepsilon}E_{\mu\nu}
=(\mathrm{sign}\,\varepsilon)\frac{E_{\mu\nu}}{|E_{\lambda}^{\lambda}|}
\end{equation}

If
\begin{equation}\label{1.5.13}
E_{\mu}^{\mu}=0
\end{equation}
then
\begin{equation}\label{1.5.14}
s(\upsilon)=0
\end{equation}
Put
\begin{equation}\label{1.5.15}
\upsilon^{0}\upsilon_{0}=1\qquad \upsilon^{i}\upsilon_{i}=-1
\end{equation}
We obtain
\begin{equation}\label{1.5.16}
\varepsilon=\frac{1}{8\pi\varkappa}E_{0}^{0}\qquad
\upsilon_{\mu}\upsilon_{\nu}=\frac{E_{\mu\nu}}{E_{0}^{0}}
\end{equation}

\subsection{The reductive synchronous reference}

Quantum jumps imply the existence of a universal (cosmic) time, so
that the spacetime manifold has the structure of the direct
product:
\begin{equation}\label{1.6.1}
M^{4}=T\times S
\end{equation}
where $T$ is cosmic time and $S$ is cosmic space. The
corresponding reference, or gauge, is reductive synchronous one
with metric
\begin{equation}\label{1.6.2}
\mathrm{d}s^{2}=\mathrm{d}t^{2}+g_{ij}\mathrm{d}x^{i}\mathrm{d}x^{j}
\qquad T\ni t=x^{0}
\end{equation}
(Metric which admits of the global clock synchronization is a
time-orthogonal one, i.e., with $g_{0i}=0$; with $g_{00}=1$, in
addition, $t$ represents the proper time at every point $s\in S$
[12].)

Now there are 11 equations (1.4.3), (1.4.2) for 11 functions
$g_{ij}\,,\;\varepsilon,\;\upsilon^{\mu}$, or, alternatively, 6
equations (1.5.3) for 6 functions $g_{ij}$; in the latter case,
the pseudomatter variables are expressed in terms of $E_{\mu\nu}$.

\subsection{Static spacetime, reference, and metric}

Static spacetime manifold is of the form
\begin{equation}\label{1.7.1}
M=T\times S
\end{equation}
and there exists a static reference, or gauge, with metric
\begin{equation}\label{1.7.2}
\mathrm{d}s^{2}=g_{00}(s)\mathrm{d}t^{2}+g_{ij}(s)\mathrm{d}x^{i}
\mathrm{d}x^{j}\qquad \mathrm{d}t=\mathrm{d}x^{0}\qquad s\in S
\end{equation}
In view of invariance with respect to $\mathrm{d}t\rightarrow-
\mathrm{d}t$, we have
\begin{equation}\label{1.7.3}
E_{0i}\equiv 0\,,\;\;i=1,2,3
\end{equation}
so that
\begin{equation}\label{1.7.4}
\upsilon_{0}\upsilon_{i}=0
\end{equation}

{\it Stress pseudomatter}:
\begin{equation}\label{1.7.5}
\upsilon_{0}=0\qquad T_{\mathrm{ps}\,0\mu}=0\qquad
T_{\mathrm{ps}\,ij}\neq 0
\end{equation}
There are 8 equations:
\begin{equation}\label{1.7.6}
E_{00}=0
\end{equation}
\begin{equation}\label{1.7.7}
E_{ij}=8\pi\varkappa\,\varepsilon\upsilon_{i}\upsilon_{j}
\end{equation}
\begin{equation}\label{1.7.8}
\upsilon^{i}\upsilon_{i}=-1\qquad s(\upsilon)=-1
\end{equation}
for 11 functions:
$g_{00},\;g_{ij}\,,\;\varepsilon,\;\upsilon_{i}$. We may preassign
$\varepsilon$ and $\upsilon_{i}$ (under condition (1.7.8)), then
there are 7 equations (1.7.6), (1.7.7) for $g_{00},\;g_{ij}$.
Alternatively, if $g$ and $T$ are given, then
\begin{equation}\label{1.7.9}
\varepsilon=-\frac{1}{8\pi\varkappa}E_{l}^{l}
\end{equation}
\begin{equation}\label{1.7.10}
\upsilon_{i}\upsilon_{j}=-\frac{E_{ij}}{E_{l}^{l}}
\end{equation}

{\it Energy (``cold'') pseudomatter}:
\begin{equation}\label{1.7.11}
\upsilon_{i}=0\qquad \upsilon^{0}\upsilon_{0}=1\qquad
s(\upsilon)=1\qquad T_{\mathrm{ps}\, i\mu}=0\qquad T_{00}\neq 0
\end{equation}
There are 7 equations:
\begin{equation}\label{1.7.12}
E_{0}^{0}=8\pi\varkappa\,\epsilon
\end{equation}
\begin{equation}\label{1.7.13}
E_{j}^{i}=0
\end{equation}
for 8 functions: $g_{00},\;g_{ij},\;\epsilon$. We may preassign
$\epsilon$; alternatively, with $g$ ant $T$ given,
\begin{equation}\label{1.7.14}
\epsilon =\frac{1}{8\pi\varkappa}E_{0}^{0}
\end{equation}

\section{Spherical static objects}

\subsection{Spherical static reference frame}

In spherical coordinates,
$x^{1}=r,\;x^{2}=\theta,\;x^{3}=\varphi$, metric is of the form
\begin{equation}\label{2.1.1}
\mathrm{d}s^{2}=\mathrm{e}^{\nu}\mathrm{d}t^{2}-\mathrm{e}^{\lambda}
\mathrm{d}r^{2}-r^{2}(\mathrm{d}\theta^{2}+\sin^{2}\theta\,\mathrm{d}\varphi^{2})\qquad
\nu=\nu(r),\;\lambda=\lambda(r)
\end{equation}
In view of spherical symmetry and invariance with respect to
$\mathrm{d}x^{2}\rightarrow-\mathrm{d}x^{2},\;\mathrm{d}x^{3}\rightarrow-\mathrm{d}x^{3}$
we have
\begin{equation}\label{2.1.2}
E_{k\mu}=E_{kk}\delta_{\mu k},\;\;k=2,3
\end{equation}
\begin{equation}\label{2.1.3}
E_{2}^{2}=E^{3}_{3}
\end{equation}
so that
\begin{equation}\label{2.1.4}
\upsilon^{2}=\upsilon^{3}=0\qquad
\upsilon^{0}=\upsilon^{0}(r)\qquad\upsilon^{1}=\upsilon^{1}(r)
\end{equation}
The equations are
\begin{equation}\label{2.1.5}
E_{0}^{0}=8\pi\varkappa\,\varepsilon\upsilon^{0}\upsilon_{0}
\end{equation}
\begin{equation}\label{2.1.6}
E_{1}^{1}=8\pi\varkappa\,\varepsilon\upsilon^{1}\upsilon_{1}
\end{equation}
\begin{equation}\label{2.1.7}
E_{2}^{2}=0
\end{equation}
\begin{equation}\label{2.1.8}
\upsilon^{0}\upsilon^{1}=0
\end{equation}
\begin{equation}\label{2.1.9}
\upsilon^{0}\upsilon_{0}+\upsilon^{1}\upsilon_{1}=s(\upsilon)
\end{equation}

{\it Stress pseudomatter}:
\begin{equation}\label{2.1.10}
\upsilon^{0}=0\qquad \upsilon^{1}\upsilon_{1}=-1\qquad
s(\upsilon)=-1
\end{equation}
There are 3 equations:
\begin{equation}\label{2.1.11}
E^{0}_{0}=0
\end{equation}
\begin{equation}\label{2.1.12}
E^{1}_{1}=-8\pi\varkappa\,\varepsilon
\end{equation}
\begin{equation}\label{2.1.13}
E^{2}_{2}=0
\end{equation}
for 3 functions:
$\mathrm{e}^{\nu},\;\mathrm{e}^{\lambda},\;\varepsilon$.

{\it Energy pseudomatter}:
\begin{equation}\label{2.1.14}
\upsilon^{1}=0\qquad \upsilon^{0}\upsilon_{0}=1\qquad
s(\upsilon)=1
\end{equation}
There are 3 equations:
\begin{equation}\label{2.1.15}
E_{0}^{0}=8\pi\varkappa\,\varepsilon
\end{equation}
\begin{equation}\label{2.1.16}
E^{1}_{1}=0
\end{equation}
\begin{equation}\label{2.1.17}
E^{2}_{2}=0
\end{equation}
for 3 functions:
$\mathrm{e}^{\nu},\;\mathrm{e}^{\lambda},\;\varepsilon$.

\subsection{The Einstein tensor. Matter distribution}

The components of the Einstein tensor $G$ are [12]
\begin{equation}\label{2.2.1}
G^{0}_{0}=\mathrm{e}^{-\lambda}\left(\frac{\lambda'}{r}-\frac{1}{r^{2}}\right)
+\frac{1}{r^{2}}
\end{equation}
\begin{equation}\label{2.2.2}
G^{1}_{1}=-\mathrm{e}^{-\lambda}\left(\frac{\nu\,'}{r}+
\frac{1}{r^{2}}\right)+\frac{1}{r^{2}}
\end{equation}
\begin{equation}\label{2.2.3}
G^{2}_{2}=-\frac{1}{2}\mathrm{e}^{-\lambda}\left( \nu\,''+
\frac{\nu\,'\,^{2}}{2}+\frac{\nu\,'}{r}-\frac{\lambda'}{r}
-\frac{\nu\,'\lambda'}{2}\right)
\end{equation}
where $':=\frac{\mathrm{d}}{\mathrm{d}r}$\,.

Let
\begin{equation}\label{2.2.4}
T=0 \quad \mathrm{for}\;r>a
\end{equation}
It is convenient to put [13]
\begin{equation}\label{2.2.5}
\mathrm{e}^{\lambda(r)}=\frac{1}{1-2m(r)/r}\qquad m(0)=0
\end{equation}
We have
\begin{equation}\label{2.2.6}
\frac{\mathrm{d}m}{\mathrm{d}r}=\frac{r^{2}}{2}G^{0}_{0}
\end{equation}
whence
\begin{equation}\label{2.2.7}
\int\limits_{0}^{a}r^{2}G^{0}_{0}\mathrm{d}r=2m(a)=:2\varkappa M=:
r_{\mathrm{S}}= r_{\mathrm{Schwarzschild}}
\end{equation}

In this Section, we put
\begin{equation}\label{2.2.8}
\Lambda=0
\end{equation}
so that
\begin{equation}\label{2.2.9}
E=G-8\pi\varkappa T
\end{equation}
Now
\begin{equation}\label{2.2.10}
G^{0}_{0}=8\pi\varkappa
(T^{0}_{0}+\varepsilon\upsilon^{0}\upsilon_{0})
\end{equation}
and we obtain
\begin{equation}\label{2.2.11}
4\pi\int\limits_{0}^{a}r^{2}(T^{0}_{0}+\varepsilon\upsilon^{0}\upsilon_{0})
\mathrm{d}r=M
\end{equation}

Let matter be a perfect fluid:
\begin{equation}\label{2.2.12}
T_{\mu}^{\nu}=(\varrho+p)v_{\mu}v^{\nu}-\delta^{\nu}_{\mu}p\qquad
v^{\mu}v_{\mu}=1
\end{equation}
In the spherical static case
\begin{equation}\label{2.2.13}
v^{i}=0\qquad v^{0}v_{0}=1
\end{equation}
so that
\begin{equation}\label{2.2.14}
T^{0}_{0}=\varrho\qquad T^{1}_{1}=T^{2}_{2}=T^{3}_{3}=-p
\end{equation}
Now
\begin{equation}\label{2.2.15}
4\pi\int\limits_{0}^{a}r^{2}(\varrho+\varepsilon \upsilon^{0}
\upsilon_{0})\mathrm{d}r=M
\end{equation}

\subsection {Dimensionless quantities}

It is convenient to introduce dimensionless quantities:
\begin{eqnarray}
x&=&\frac{r}{a}\qquad
'=\frac{\mathrm{d}}{\mathrm{d}x}\qquad\bar{r}_{\mathrm{S}}=
\frac{r_{\mathrm{S}}}{a}=\frac{2\varkappa M}{a}\nonumber\\
\bar{G}&=&a^{2}G\qquad \bar{T}=\varkappa\, a^{2}T\qquad
\bar{\varrho}=\varkappa\, a^{2}\varrho\qquad \bar{p}=\varkappa\,
a^{2}p\qquad \bar{\varepsilon} =\varkappa\, a^{2}\varepsilon
\end{eqnarray}
Introduce
\begin{equation}\label{2.3.2}
\mu=\nu\,'\qquad
\nu(x)=\nu(1)-\int\limits_{x}^{1}\mu(x)\mathrm{d}x\qquad\nu\,''=\mu'
\end{equation}
Now
\begin{equation}\label{2.3.3}
\bar{G}^{0}_{0}=\frac{\lambda'\mathrm{e}^{-\lambda}}{x}
+\frac{1-\mathrm{e}^{-\lambda}}{x^{2}}
\end{equation}
\begin{equation}\label{2.3.4}
\bar{G}^{1}_{1}=-\frac{\mu\mathrm{e}^{-\lambda}}{x}+
\frac{1-\mathrm{e}^{-\lambda}}{x^{2}}
\end{equation}
\begin{equation}\label{2.3.5}
\bar{G}^{2}_{2}=-\frac{1}{2}\mathrm{e}^{-\lambda}\left[\mu' +
\left(\frac{1}{x}+\frac{\mu}{2}\right)\left(\mu-\lambda'\right)
\right]
\end{equation}
(2.3.3)--(2.3.5) imply helpful formulas [14]:
\begin{equation}\label{2.3.6}
\mathrm{e}^{-\lambda}=1-\frac{1}{x}\int\limits_{0}^{x}x^{2}\bar{G}^{0}_{0}
\mathrm{d}x
\end{equation}
\begin{equation}\label{2.3.7}
\mu=\mathrm{e}^{\lambda}x\left(\bar{G}^{0}_{0}-\bar{G}^{1}_{1}\right)-
\lambda'
\end{equation}
\begin{equation}\label{2.3.8}
\bar{G}^{2}_{2}=\frac{1}{2}x\left(\bar{G}^{1}_{1}\right)'+\left(1+
\frac{1}{4}x\mu\right)\bar{G}^{1}_{1}-\frac{1}{4}x\mu\bar{G}^{0}_{0}
\end{equation}
(2.2.7) amounts to
\begin{equation}\label{2.3.9}
\int\limits_{0}^{1}x^{2}\bar{G}^{0}_{0}\mathrm{d}x=\bar{r}_{\mathrm{S}}
\end{equation}
The equations are:
\begin{equation}\label{2.3.10}
\bar{G}^{0}_{0}=8\pi\{\bar{T}^{0}_{0}+\theta[s(\upsilon)]
\bar{\varepsilon}\}
\end{equation}
\begin{equation}\label{2.3.11}
\bar{G}^{1}_{1}=8\pi\{\bar{T}^{1}_{1}-\theta[-s(\upsilon)]
\bar{\varepsilon}\}
\end{equation}
\begin{equation}\label{2.3.12}
\bar{G}^{2}_{2}=8\pi\bar{T}^{2}_{2}
\end{equation}
where $\theta$ is the step-function,
\begin{equation}\label{2.3.13}
s(\upsilon)=1/-1 \quad\mathrm{for \;energy/stress\;pseudomatter}
\end{equation}

For perfect fluid, the equations are:
\begin{equation}\label{2.3.14}
\bar{G}^{0}_{0}=8\pi\{\bar{\varrho}+\theta[s(\upsilon)]
\bar{\varepsilon}\}
\end{equation}
\begin{equation}\label{2.3.15}
\bar{G}^{1}_{1}=-8\pi\{\bar{p}+\theta[-s(\upsilon)]
\bar{\varepsilon}\}
\end{equation}
\begin{equation}\label{2.3.16}
\bar{G}^{2}_{2}=-8\pi\bar{p}
\end{equation}

\subsection{Nonexistence of matterless pseudomatter static objects}

Without matter, the equations are:
\begin{equation}\label{2.4.1}
\bar{G}^{0}_{0}=8\pi\theta[s(\upsilon)] \bar{\varepsilon}
\end{equation}
\begin{equation}\label{2.4.2}
\bar{G}^{1}_{1}=-8\pi\theta[-s(\upsilon)] \bar{\varepsilon}
\end{equation}
\begin{equation}\label{2.4.3}
\bar{G}^{2}_{2}=0
\end{equation}

Let
\begin{equation}\label{2.4.4}
s(\upsilon)=1
\end{equation}
then $\bar{G}^{1}_{1}=0$, and from (2.3.8) follows
\begin{equation}\label{2.4.5}
\mu\bar{G}^{0}_{0}=0
\end{equation}
We have
\begin{equation}\label{2.4.6}
\mu=0\Rightarrow
\;[\mathrm{by}\;(2.3.4)]\;\lambda=0\;\Rightarrow\;
[\mathrm{by}\;(2.3.3)]\;\bar{G}^{0}_{0}=0
\end{equation}
Thus $\bar{G}^{0}_{0}=0$, so that, by (2.4.1),
\begin{equation}\label{2.4.7}
\bar{\varepsilon}=0
\end{equation}

Now let
\begin{equation}\label{2.4.8}
s(\upsilon)=-1
\end{equation}
then
\begin{equation}\label{2.4.9}
\bar{G}^{0}_{0}=0\;\Rightarrow
\;[\mathrm{by}\;(2.3.6)]\;\lambda=0\;\Rightarrow\;
[\mathrm{by}\;(2.3.4)]\;\mu=-x\bar{G}^{1}_{1}
\end{equation}
and (2.3.8) amounts to
\begin{equation}\label{2.4.10}
\frac{1}{2}x(\bar{G}^{1}_{1})'+\bar{G}^{1}_{1}-
\frac{1}{4}x^{2}(\bar{G}^{1}_{1})^{2}=0
\end{equation}
The solution is
\begin{equation}\label{2.4.11}
\bar{G}^{1}_{1}(x)=\frac{1}{x^{2}[1/\bar{G}^{1}_{1}(1)-(1/2)\ln
x]}
\end{equation}
If $\bar{G}^{1}_{1}(1)\neq 0$, then $\bar{G}^{1}_{1}(x)$ is
singular at $x=0$. Thus $\bar{G}^{1}_{1}(1)=0$, so that, by
(2.4.2),
\begin{equation}\label{2.4.12}
\bar{\varepsilon}=0
\end{equation}

With matter present, energy pseudomatter may only exist where
$T\neq 0$.

It appears that, in view of (1.7.4), the results of this
Subsection are valid in an arbitrary static case.

\subsection{The Schwarzschild solution with energy pseudomatter}

Let there be a perfect fluid and energy pseudomatter for $0\leq
x\leq 1$ (interior) and vacuum for $x>1$ (exterior). The interior
equations (2.3.14)--(2.3.16) are
\begin{equation}\label{2.5.1}
\bar{G}^{0}_{0}=8\pi\bar{\varrho}_{\mathrm{eff}}
\end{equation}
\begin{equation}\label{2.5.2}
\bar{G}^{1}_{1}=-8\pi\bar{p}
\end{equation}
\begin{equation}\label{2.5.3}
\bar{G}^{2}_{2}=-8\pi\bar{p}
\end{equation}
where
\begin{equation}\label{2.5.4}
\bar{\varrho}_{\mathrm{effective}}=:\bar{\varrho}_{\mathrm{eff}}=
\bar{\varrho}+\bar{\varepsilon}
\end{equation}
and
\begin{equation}\label{2.5.5}
8\pi\int\limits_{0}^{1}x^{2}\bar{\varrho}_{\mathrm{eff}}\mathrm{d}x=
\bar{r}_{\mathrm{S}}
\end{equation}
The exterior equations are (2.5.1)--(2.5.3) with
$\bar{\varrho}=\bar{\varepsilon}=0,\;\bar{p}=0$.

Let us consider the Schwarzschild solution to the above equations.
We have [11,13,14]:
\begin{equation}\label{2.5.6}
\bar{\varrho}_{\mathrm{eff}}=\mathrm{const}=\frac{3}{8\pi}\bar{r}_{\mathrm{S}}
\end{equation}
\begin{equation}\label{2.5.7}
\bar{r}_{\mathrm{S}}<\frac{8}{9}
\end{equation}
\begin{equation}\label{2.5.8}
\bar{p}=\bar{\varrho}_{\mathrm{eff}}
\frac{(1-\bar{r}_{\mathrm{S}}x^{2})^{1/2}-(1-\bar{r}_{\mathrm{S}})^{1/2}}
{3(1-\bar{r}_{\mathrm{S}})^{1/2}-(1-\bar{r}_{\mathrm{S}}x^{2})^{1/2}}
\end{equation}

Let $p$ satisfy the standard condition:
\begin{equation}\label{2.5.9}
\bar{p}\leq\frac{1}{3}\bar{\varrho}_{\mathrm{eff}}
\end{equation}
If
\begin{equation}\label{2.5.10}
\bar{p}>\frac{1}{3}\bar{\varrho}_{\mathrm{eff}}
\end{equation}
then
\begin{equation}\label{2.5.11}
\frac{1}{3}(\bar{\varrho}+\bar{\varepsilon})
<\bar{p}\leq\frac{1}{3}\bar{\varrho}
\end{equation}
whence
\begin{equation}\label{2.5.12}
\bar{\varepsilon}<0
\end{equation}
From (2.5.10), (2.5.8) follows
\begin{equation}\label{2.5.13}
\bar{r}_{\mathrm{S}}>\frac{5}{9}\quad\mathrm{and}\quad
x<\frac{3}{2\bar{r}_{\mathrm{S}}^{1/2}}\left(\bar{r}_{\mathrm{S}}-
\frac{5}{9}\right)^{1/2}\;\Rightarrow\;\bar{\varepsilon}<0
\end{equation}
Specifically,
\begin{equation}\label{2.5.14}
\mathrm{for}\;\bar{r}_{\mathrm{S}}\rightarrow\frac{8}{9}\qquad
x<\left(\frac{27}{32}\right)^{1/2}
\end{equation}
Thus the realistic condition (2.5.3) implies the existence of
energy pseudomatter with negative density.

\subsection{Energy pseudomatter halo}

Let
\begin{equation}\label{2.6.1}
\bar{r}_{\mathrm{S}}\ll 1\qquad
\bar{\varrho}_{\mathrm{eff}}=\frac{3}{8\pi}\bar{r}_{\mathrm{S}}\ll
1
\end{equation}
then
\begin{equation}\label{2.6.2}
\bar{p}\approx
\frac{1}{4}\bar{\varrho}_{\mathrm{eff}}\bar{r}_{\mathrm{S}}(1-x^{2})
=\frac{2\pi}{3}\bar{\varrho}_{\mathrm{eff}}^{2}(1-x^{2})
\ll\bar{\varrho}_{\mathrm{eff}}
\end{equation}
Introduce
\begin{equation}\label{2.6.3}
0<x_{\mathrm{m}}\ll 1\qquad x_{\mathrm{m}}:=x_{\mathrm{matter}}
\end{equation}
and put
\begin{equation}\label{2.6.4}
\mathrm{for}\;\;0\leq x\lesssim x_{\mathrm{m}}\qquad \bar{\varrho}
\approx\bar{\varrho}_{\mathrm{eff}}\gg\bar{\varepsilon}\approx 0
\end{equation}
\begin{equation}\label{2.6.5}
\mathrm{for}\;\;x_{\mathrm{m}}\lesssim x\leq
1\qquad\bar{\varrho}\approx
3\bar{p}=2\pi\bar{\varrho}_{\mathrm{eff}}^{2}(1-x^{2})
\ll\bar{\varepsilon}\approx\bar{\varrho}_{\mathrm{eff}}
\end{equation}
This is a material object in the small interior (2.6.4) and an
energy pseudomatter halo with little matter in the large exterior
(2.6.5).

In the halo,
\begin{equation}\label{2.6.6}
\bar{M}(x)=4\pi\int\limits_{0}^{x}x^{2}
\bar{\varrho}_{\mathrm{eff}}\mathrm{d}x\propto x^{3}
\end{equation}
so that we obtain for the velocity $v$:
\begin{equation}\label{2.6.7}
\frac{v^{2}}{x}\propto\frac{\bar{M}(x)}{x^{2}}\propto x\qquad
v\propto x
\end{equation}

\subsection{Stress pseudomatter halo}

Let us consider the case when in the exterior, $x>1$, there is
only stress pseudomatter. The exterior equations are:
\begin{equation}\label{2.7.1}
\bar{G}^{0}_{0}=0
\end{equation}
\begin{equation}\label{2.7.2}
\bar{G}^{1}_{1}=-8\pi\bar{\varepsilon}
\end{equation}
\begin{equation}\label{2.7.3}
\bar{G}^{2}_{2}=0
\end{equation}
In the exterior, we obtain from (2.3.8), (2.3.4)
\begin{equation}\label{2.7.4}
(\bar{G}^{1}_{1})'+\frac{2}{x}\left[1+
\frac{1}{4}(\mathrm{e}^{\lambda}-1)\right]\bar{G}^{1}_{1}
-\frac{1}{2}x\mathrm{e}^{\lambda}(\bar{G}^{1}_{1})^{2}=0
\end{equation}
and from (2.3.6), (2.3.9)
\begin{equation}\label{2.7.5}
\mathrm{e}^{\lambda}=\frac{1}{1-\bar{r}_{\mathrm{S}}/x}
\end{equation}
Thus we obtain the equation for $\bar{G}^{1}_{1}$:
\begin{equation}\label{2.7.6}
(\bar{G}^{1}_{1})'+\frac{2}{x}\left[1+
\frac{1}{4}\frac{\bar{r}_{\mathrm{S}}}{x-
\bar{r}_{\mathrm{S}}}\right]\bar{G}^{1}_{1} -\frac{1}{2}
\frac{x^{2}}{x-\bar{r}_{\mathrm{S}}}(\bar{G}^{1}_{1})^{2}=0
\end{equation}
Put
\begin{equation}\label{2.7.7}
\bar{G}^{1}_{1}=\frac{1}{y}
\end{equation}
which gives
\begin{equation}\label{2.7.8}
y'-\frac{2}{x}\left[1+ \frac{1}{4}\frac{\bar{r}_{\mathrm{S}}}{x-
\bar{r}_{\mathrm{S}}}\right]y=-\frac{1}{2}
\frac{x^{2}}{x-\bar{r}_{\mathrm{S}}}
\end{equation}
whence
\begin{eqnarray}
y=x^{3/2}\left(\frac{x-\bar{r}_{\mathrm{S}}}
{1-\bar{r}_{\mathrm{S}}}\right)^{1/2}\hspace{120mm}\nonumber\\\times\left\{y(1)-
\frac{1}{2}(1-\bar{r}_{\mathrm{S}})^{1/2}\left[\ln
x+2\ln\frac{1+\sqrt{1-\bar{r}_{\mathrm{S}}/x}}
{1+\sqrt{1-\bar{r}_{\mathrm{S}}}}+2\left(\frac{1}
{\sqrt{1-\bar{r}_{\mathrm{S}}}}-\frac{1}
{\sqrt{1-\bar{r}_{\mathrm{S}}/x}}\right)\right]\right\}\;\;\,
\end{eqnarray}

To obtain metric (2.1.1), we have to find $\mathrm{e}^{\nu}$. From
(2.3.2) follows
\begin{equation}\label{2.7.10}
\mathrm{e}^{\nu}=[\mathrm{e}^{\nu}](1)\exp\left\{
\int\limits_{1}^{x}\mu\,\mathrm{d}x\right\}
\end{equation}
and from (2.3.4)
\begin{equation}\label{2.7.11}
\mu=\frac{\mathrm{e}^{\lambda}-1}{x}-x\mathrm{e}^{\lambda}\bar{G}^{1}_{1}
=\frac{\bar{r}_{\mathrm{S}}}{x(x-\bar{r}_{\mathrm{S}})}-
\frac{x^{2}}{(x-\bar{r}_{\mathrm{S}})y}
\end{equation}
so that
\begin{equation}\label{2.7.12}
\mathrm{e}^{\nu}=[\mathrm{e}^{\nu}](1)
\frac{1}{1-\bar{r}_{\mathrm{S}}}
\left(1-\frac{\bar{r}_{\mathrm{S}}}{x}\right)\exp\left\{-\int\limits_{1}^{x}
\frac{x^{2}}{(x-\bar{r}_{\mathrm{S}})y}\mathrm{d}x\right\}
\end{equation}

$[\mathrm{e}^{\nu}](1)$ depends on the choice of $t$ in (2.1.1).
In view of (2.7.5), we introduce the condition
\begin{equation}\label{2.7.13}
\left[\frac{\mathrm{d}}{\mathrm{d}x}\mathrm{e}^{\nu}\right](1)
=\left[\frac{\mathrm{d}}{\mathrm{d}x}\left(c+1-
\frac{\bar{r}_{\mathrm{S}}}{x}\right)\right](1)
\end{equation}
i.e.,
\begin{equation}\label{2.7.14}
\left[\frac{\mathrm{d}}{\mathrm{d}x}\mathrm{e}^{\nu}\right](1)
=\bar{r}_{\mathrm{S}}
\end{equation}

In what follows in this Subsection, we consider the case
\begin{equation}\label{2.7.15}
\bar{r}_{\mathrm{S}}\ll 1
\end{equation}
From (2.7.9) follows
\begin{equation}\label{2.7.16}
y\approx x^{2}\left[y(1)-\frac{1}{2}\ln x\right]
\end{equation}
so that to avoid a singularity, we put
\begin{equation}\label{2.7.17}
y(1)=-b\qquad b>0
\end{equation}
Thus
\begin{eqnarray}
\mathrm{e}^{\nu}&=&[\mathrm{e}^{\nu}](1)\frac{1}{1-\bar{r}_{\mathrm{S}}}
\left(1-\frac{\bar{r}_{\mathrm{S}}}{x}\right)\left(1+\frac{1}
{2b}\ln x\right)^{2}\nonumber\\ &\approx&
[\mathrm{e}^{\nu}](1)\frac{1}{1-\bar{r}_{\mathrm{S}}}
\left[1-\frac{\bar{r}_{\mathrm{S}}}{x} +\frac{\ln x}{b}+\frac{(\ln
x)^{2}}{4b^{2}}\right]
\end{eqnarray}

Now put
\begin{equation}\label{2.7.19}
\mathrm{e}^{\nu}=g_{00}=1+2\varphi\qquad \varphi\ll 1
\end{equation}
where $\varphi$ is the Newtonian potential [12,15], and
\begin{equation}\label{2.7.20}
[\mathrm{e}^{\nu}](1)\frac{1}{1-\bar{r}_{\mathrm{S}}}= 1+\beta
\qquad|\beta|\ll 1
\end{equation}
Then
\begin{equation}\label{2.7.21}
\varphi=\frac{\beta}{2}+\frac{1+\beta}{2}\left[-
\frac{\bar{r}_{\mathrm{S}}}{x}+\frac{\ln x}{b}+ \frac{(\ln x
)^{2}}{4b^{2}}\right]
\end{equation}
and
\begin{equation}\label{2.7.22}
\varphi\,'=\frac{1+\beta}{2}\left[
\frac{\bar{r}_{\mathrm{S}}}{x^{2}}+\frac{1}{bx}+ \frac{\ln x
}{2b^{2}x}\right]
\end{equation}
From (2.7.14) and (2.7.19) follows
\begin{equation}\label{2.7.23}
\varphi\,'(1)=\frac{\bar{r}_{\mathrm{S}}}{2}
\end{equation}
and
\begin{equation}\label{2.7.24}
\beta=-\frac{1}{1+\bar{r}_{\mathrm{S}}b}\qquad 1+\beta=
\frac{1}{1+1/\bar{r}_{\mathrm{S}}b}\qquad \bar{r}_{\mathrm{S}}b\gg
1
\end{equation}
Next,
\begin{equation}\label{2.7.25}
\ln x \ll b
\end{equation}
so that
\begin{equation}\label{2.7.26}
\varphi\,'=\frac{1}{2(1+1/\bar{r}_{\mathrm{S}}b)}
\left(\frac{\bar{r}_{\mathrm{S}}}{x^{2}}+\frac{1}{bx}\right)
\end{equation}
We have for the velocity $v$
\begin{equation}\label{2.7.27}
\frac{v^{2}}{x}=\varphi\,' \qquad v\ll 1
\end{equation}
whence
\begin{equation}\label{2.2.28}
v^{2}=\frac{1}{2(1+1/\bar{r}_{\mathrm{S}}b)}
\left(\frac{\bar{r}_{\mathrm{S}}}{x}+\frac{1}{b}\right)
\end{equation}
Specifically,
\begin{equation}\label{2.7.29}
\mathrm{for}\quad 1\ll\bar{r}_{\mathrm{S}}b\ll x\ll \mathrm{e}^{b}
\qquad v=\sqrt{\frac{1}{2b}}=\mathrm{const}
\end{equation}
This result is in good agreement with galaxy-scale observations
[6].

It might be worthwhile to point out that in the treatment in this
Subsection, the localization of the matter and gravitational
potential are spatially separated; this takes place in the bullet
cluster$~\,$[6].

\section{Dynamical objects}

\subsection{Collapse with pseudomatter}

In the treatment of a collapse, we will exploit the relevant
results of [11]. In the spherical case, with
$x^{1}=r,\;x^{2}=\theta,\;x^{3}=\varphi$, the metric is of the
form
\begin{equation}\label{3.1.1}
\mathrm{d}s^{2}=\mathrm{d}t^{2}-U(r,t)\mathrm{d}r^{2}
-V(r,t)(\mathrm{d}\theta^{2}+\sin^{2}\theta\,\mathrm{d}\varphi^{2})
\end{equation}
The quantities which are not identically zero are:
\begin{eqnarray}
G_{00}\qquad G_{01}\qquad G_{11}\qquad G_{22}\qquad
G_{3}^{3}=G_{2}^{2}\nonumber\\ E_{00}\qquad E_{01}\qquad
E_{11}\qquad E_{22}\qquad E_{3}^{3}=E_{2}^{2}
\end{eqnarray}
Now we use the metrodynamical equation (1.5.3) with $E_{00}\neq
0$:
\begin{equation}\label{3.1.3}
E_{ij}=\frac{E_{0i}E_{0j}}{E_{00}}
\end{equation}
whence
\begin{equation}\label{3.1.4}
E_{11}=\frac{(E_{01})^{2}}{E_{00}}
\end{equation}
\begin{equation}\label{3.1.5}
E_{22}=0
\end{equation}
We put
\begin{equation}\label{3.1.6}
\Lambda=0\qquad E=G-8\pi\varkappa T
\end{equation}
so that
\begin{equation}\label{3.1.7}
G_{22}-8\pi\varkappa T_{22}=0
\end{equation}
\begin{equation}\label{3.1.8}
G_{11}-8\pi\varkappa T_{11}=\frac{(G_{01}-8\pi\varkappa
T_{01})^{2}}{G_{00}-8\pi\varkappa T_{00}}
\end{equation}

Now we consider the simplest case: a solution to (3.1.7), (3.1.8)
for which
\begin{equation}\label{3.1.9}
\mathrm{d}s^{2}=\mathrm{d}t^{2}-
R^{2}(t)\left[\frac{\mathrm{d}r^{2}}{1-kr^{2}}+
r^{2}(\mathrm{d}\theta^{2}+\sin^{2}\theta\,\mathrm{d}\varphi^{2})\right]
\end{equation}
\begin{equation}\label{3.1.10}
G_{22}=0\qquad G_{11}=0\qquad G_{01}=0\qquad
G_{00}=8\pi\varkappa\,\sigma\qquad
\sigma=\frac{\sigma(0)}{R^{3}(t)}
\end{equation}
\begin{equation}\label{3.1.11}
k=\frac{8\pi\varkappa}{3}\sigma(0)=\frac{2\varkappa M}{a^{3}}
\qquad ka^{2}<1\qquad \frac{2\varkappa M}{a}<1
\end{equation}
$R(t)$ is given in [11]. Again,
\begin{equation}\label{3.1.12}
T_{22}=0\qquad T_{2}^{2}=0
\end{equation}
\begin{equation}\label{3.1.13}
T_{11}=\frac{(T_{01})^{2}}{T_{00}-\sigma}\qquad\qquad
-\frac{R^{2}}{1-kr^{2}}T_{1}^{1}=\frac{(T_{1}^{0})^{2}}{T_{0}^{0}-\sigma}
\end{equation}

Pseudomatter is determined by (1.4.3), (1.4.2):
\begin{equation}\label{3.1.14}
E_{\mu}^{\nu}=8\pi\varkappa\,\varepsilon\upsilon_{\mu}\upsilon^{\nu}
\qquad\upsilon_{\mu}\upsilon^{\mu}=s(\upsilon)
\end{equation}
From (3.1.5) follows
\begin{equation}\label{3.1.15}
\upsilon_{2}=\upsilon_{3}=0
\end{equation}
Thus
\begin{equation}\label{3.1.16}
\upsilon_{0}\upsilon^{0}+\upsilon_{1}\upsilon^{1}=s(\upsilon)
\end{equation}
\begin{equation}\label{3.1.17}
\varepsilon\upsilon_{0}\upsilon^{0}=\sigma-T_{0}^{0}\qquad
\varepsilon\upsilon_{1}\upsilon^{1}=-T_{1}^{1}
\end{equation}
whence
\begin{equation}\label{3.1.18}
\varepsilon s(\upsilon)=\sigma-T_{0}^{0}-T^{1}_{1}
\end{equation}
Next,
\begin{equation}\label{3.1.19}
T_{0}^{0}=\varrho \qquad T_{1}^{1}-p_{r}
\end{equation}
where $p_{r}$ is the radial pressure. Thus
\begin{equation}\label{3.1.20}
\frac{R^{2}}{1-kr^{2}}p_{r}=\frac{(T_{1}^{0})^{2}}{\varrho-\sigma}
\end{equation}
\begin{equation}\label{3.1.21}
\varepsilon\upsilon_{0}\upsilon^{0}=
\sigma-\varrho\qquad\varepsilon\upsilon_{1}\upsilon^{1}=p_{r}
\end{equation}
\begin{equation}\label{3.1.22}
-\varepsilon s(\upsilon)=D
\end{equation}
where
\begin{equation}\label{3.2.23}
D=\varrho-p_{r}-\sigma
\end{equation}
We have
\begin{equation}\label{3.1.24}
p_{r}>0\qquad \varrho-\sigma>0
\end{equation}
so that
\begin{equation}\label{3.1.25}
\varepsilon<0
\end{equation}
We find
\begin{equation}\label{3.1.26}
s(\upsilon)=\mathrm{sign}\,D
\end{equation}

If
\begin{equation}\label{3.1.27}
D\neq 0
\end{equation}
then
\begin{equation}\label{3.1.28}
\varepsilon =-|D|
\end{equation}
\begin{equation}\label{3.1.29}
\upsilon_{0}\upsilon^{0}=\frac{\varrho-\sigma}{|D|} \qquad
\upsilon_{1}\upsilon^{1}=-\frac{p_{r}}{|D|}
\end{equation}

If
\begin{equation}\label{3.1.30}
D=0
\end{equation}
then
\begin{equation}\label{3.1.31}
s(\upsilon)=0\qquad\upsilon_{0}\upsilon^{0}+
\upsilon_{1}\upsilon^{1}=0
\end{equation}
Put
\begin{equation}\label{3.1.32}
\upsilon_{0}\upsilon^{0}=1\qquad\upsilon_{1}\upsilon^{1}=-1
\end{equation}
then
\begin{equation}\label{3.1.33}
\varepsilon=-(\varrho-\sigma)
\end{equation}

Let
\begin{equation}\label{3.1.34}
R\rightarrow 0
\end{equation}
We have
\begin{equation}\label{3.1.35}
\varrho\propto\frac{1}{R^{4}}\qquad p_{r}\propto\frac{1}{R^{4}}
\qquad T_{1}^{0}\propto\frac{1}{R^{3}}
\end{equation}
let $p_{r}<\varrho$, then
\begin{equation}\label{3.1.36}
D>0\qquad D\propto\frac{1}{R^{4}}\qquad
s(\upsilon)=1\qquad\upsilon_{0}\upsilon^{0}=\frac{\varrho}{\varrho-p_{r}}>1
\qquad\upsilon_{1}\upsilon^{1}=-\frac{p_{r}}{\varrho-p_{r}}
\end{equation}

\subsection{The FLRW universe with energy (``cold'') pseudomatter}

The FLRW universe metric is of the form [16]
\begin{equation}\label{3.2.1}
\mathrm{d}s^{2}=\mathrm{d}t^{2}-
R^{2}(t)\left[\frac{\mathrm{d}r^{2}}{1-kr^{2}}
+r^{2}(\mathrm{d}\theta^{2}+\sin^{2}\mathrm{d}\varphi^{2})\right],
\quad k=1,0,-1
\end{equation}
$(t=x^{0},\;r=x^{1},\;\theta=x^{2},\;\varphi=x^{3})$. The
dynamical Einstein tensor (1.1.1) is
\begin{equation}\label{3.2.2}
E=G-\Lambda g-8\pi\varkappa T
\end{equation}
and the metrodynamical equation (1.5.3) is
\begin{equation}\label{3.2.3}
E_{00}E_{ij}-E_{0i}E_{0j}=0
\end{equation}
From symmetry considerations follows
\begin{equation}\label{3.2.4}
E_{0i}\equiv 0\qquad E_{ij}\equiv 0 \;\;\mathrm{for}\;\,j\neq i
\end{equation}
Thus (3.2.3) reduces to
\begin{equation}\label{3.2.5}
E_{00}E_{ii}=0
\end{equation}

Now
\begin{equation}\label{3.2.6}
E_{\mu\nu}=8\pi\varkappa\,\upsilon_{\mu}\upsilon_{\nu}\qquad
\upsilon_{\mu}\upsilon^{\mu}=s(\upsilon)
\end{equation}
and from symmetry considerations follows
\begin{equation}\label{3.2.7}
\upsilon_{i}=0\qquad s(\upsilon)=1\qquad (\upsilon_{0})^{2}=
\upsilon_{0}\upsilon^{0}=1\qquad\mathrm{energy\;(``cold")\;pseudomatter}
\end{equation}
so that
\begin{equation}\label{3.2.8}
E_{ii}=0\qquad E_{11}=0\Leftrightarrow E_{22}=0\Leftrightarrow
E_{33}=0
\end{equation}
Thus we have two equations:
\begin{equation}\label{3.2.9}
E_{1}^{1}=0\qquad G_{1}^{1}-\Lambda-8\pi\varkappa T_{1}^{1}=0
\end{equation}
\begin{equation}\label{3.2.10}
E_{0}^{0}=8\pi\varkappa\,\varepsilon\qquad
G_{0}^{0}-\Lambda-8\pi\varkappa (T_{0}^{0}+\varepsilon)=0
\end{equation}
The components of the Einstein tensor $G$ are [16]:
\begin{equation}\label{3.2.11}
G_{0}^{0}=3\frac{\dot{R}^{2}+k}{R^{2}}\qquad
G^{1}_{1}=\frac{1}{R^{2}}(2R\ddot{R}+\dot{R}^{2}+k)
\end{equation}

Now we regard matter as a perfect fluid (2.2.12) with $v^{i}=0$
[12,16], so that
\begin{equation}\label{3.2.12}
T^{0}_{0}=\varrho\qquad T_{1}^{1}=-p
\end{equation}
and we obtain two equations:
\begin{equation}\label{3.2.13}
2R\ddot{R}+\dot{R}^{2}+k-\Lambda R^{2}+8\pi\varkappa pR^{2}=0
\end{equation}
\begin{equation}\label{3.2.14}
3(\dot{R}^{2}+k)-\Lambda
R^{2}-8\pi\varkappa(\varrho+\varepsilon)R^{2}=0
\end{equation}
for two functions: $R,\;\varepsilon$ ($p$ and $\varrho$ are
determined by (1.1.2), (1.1.6)).

From (3.2.13), (3.2.14) the energy equation follows:
\begin{equation}\label{3.2.15}
\frac{\mathrm{d}}{\mathrm{d}t}[(\varrho+\varepsilon)R^{3}]=
-p\frac{\mathrm{d}R^{3}}{\mathrm{d}t}
\end{equation}
Specifically, in the case of $k=1$ (closed universe),
$V=2\pi^{2}R^{3}$, and
\begin{equation}\label{3.2.16}
\frac{\mathrm{d}}{\mathrm{d}t}[(\varrho+\varepsilon)V]=
-p\frac{\mathrm{d}V}{\mathrm{d}t}
\end{equation}

Now, the equation
\begin{equation}\label{3.2.17}
T_{0}{}^{\mu}{}_{;\mu}=0
\end{equation}
for perfect fluid results in
\begin{equation}\label{3.2.18}
\frac{\mathrm{d}}{\mathrm{d}t}[\varrho R^{3}]=
-p\frac{\mathrm{d}R^{3}}{\mathrm{d}t}
\end{equation}
Thus
\begin{equation}\label{3.2.19}
\frac{\mathrm{d}}{\mathrm{d}t}[\varepsilon R^{3}]=0
\end{equation}
and
\begin{equation}\label{3.2.20}
\varepsilon=\frac{B}{R^{3}}\qquad B=\mathrm{const}
\end{equation}
In the actual universe $\varepsilon>0$, so that
\begin{equation}\label{3.2.21}
B>0
\end{equation}

Let
\begin{equation}\label{3.2.22}
R\rightarrow +0
\end{equation}
then
\begin{equation}\label{3.2.23}
\varrho=\frac{A}{R^{4}}\qquad
p=\frac{1}{3}\varrho=\frac{A}{3R^{4}}
\end{equation}
We have
\begin{equation}\label{3.2.24}
\frac{\varepsilon}{\varrho}=\frac{B}{A}R\qquad
\frac{\mathrm{d}}{\mathrm{d}R}\left(\frac{\varepsilon}{\varrho}\right)
=\frac{B}{A}>0
\end{equation}

Now,
\begin{equation}\label{3.2.25}
\mathrm{for}\;\,R\rightarrow\infty\qquad \varrho\propto
\frac{1}{R^{3}}\qquad\frac{\varepsilon}{\varrho}=\mathrm{const}
\end{equation}
Thus for small $R$, $\varepsilon/\varrho$ increases with time.

\section*{Acknowledgments}

I would like to thank Alex A. Lisyansky for support and Stefan V.
Mashkevich for helpful discussions.

\end{document}